\documentclass[12pt]{article}
\usepackage{a4}
\usepackage{amsfonts}
\usepackage{graphicx}

\newcommand{\RR}{\mathbb{R}}
\newcommand{\EE}{\mathbb{E}}
\newcommand{\PP}{\mathbb{P}}

\begin{document}


\title{Anticipating Activity in Social Media Spikes\footnote{This manuscript appears as Technical Report 5, 2014,
Department of Mathematics and Statistics, University of Strathclyde, UK}}

\author{Desmond J. Higham\thanks{Department of Mathematics and Statistics, University of Strathclyde, Glasgow, G1 1XH, UK},
Peter Grindrod\thanks{Mathematical Institute,
University of Oxford,
Oxford OX2 6GG, UK},
Alexander V. Mantzaris\thanks
{Department of Mathematics and Statistics, University of Strathclyde, Glasgow, G1 1XH, UK} ,
Amanda Otley\thanks{Bloom Agency,
Marshall’s Mill,
Leeds, LS11 9YJ, UK}
\and
Peter Laflin\thanks{Bloom Agency,
Marshall’s Mill,
Leeds, LS11 9YJ, UK}
}


\maketitle 


\begin{abstract}
We propose a novel mathematical model for the activity of
  microbloggers during an external, event-driven spike.
  The model leads to a testable prediction of who would become most active if a spike were 
  to take place.
  This type of information is of great interest
  to commercial organisations, governments and charities, as
  it identifies key players who can be targeted with information
  in real time when the network is most receptive.
  The model takes account of the fact that dynamic interactions
  evolve over an underlying, static network that records
  \lq\lq who listens to whom.\rq\rq\
  The model is based on the assumption that, in the case where the entire community has
  become aware of an external news event, a key driver of
  activity is the motivation to participate by responding to incoming messages.
  We test the model on a large scale Twitter conversation concerning
  the appointment of a UK Premier League football club manager. We also present
  further results for a Bundesliga football match, a marketing event and a television programme.
  In each case we find that exploiting the underlying connectivity structure
  improves the prediction of who will be active during a spike.
  We also show how the half-life of a spike in activity can be quantified in terms of the
  network size and the typical response rate.
\end{abstract}



\smallskip

\noindent
{\small
\textbf{Lay Summary:}
\textsf{
Microblogging data offers us the opportunity to understand, and
exploit, on-line human behavior.
Here, we focus on the task of predicting which users will be influential in the event of
an externally-driven spike, such as an unexpected news item.
Based on the testable hypothesis that microbloggers generate new content in 
response to
incoming messages, we develop a mathematical model and computational algorithm.
The new algorithm combines information about the static user-follower network and
the dynamic interaction patterns.
We give four Twitter case studies to show that this approach improves our ability to
anticipate
who would be most
active during a 
spike. This provides a tool for rapid targetting
of key users
when the network is at its most receptive. 
}
}

\smallskip
\smallskip

\section{Introduction}
\label{sec:intro}
Digital footprints left by our
online interactions provide a wealth of information for
social scientists
and present many new challenges in modelling and computation
 \cite{Lazer09}.
In addition to aiding our understanding of how humans interact
and make decisions
\cite{Aral2012}, microblogging data offers the prospect of
predicting future behavior
\cite{Cu2012}
and
engaging in targeted intervention
\cite{Poked}.
Commercial organisations, governments and charities are now able to interact with
the general public during the course of an online, global conversation, and exploit opportunities to leverage current sentiment.
We focus here on the specific case where a rapid spike of activity can be attributed to a high profile event or news item. (For example, a pivotal moment  in a sporting event,
or an unexpected reality TV voting result.)
In Figure~\ref{fig:spikes} we show examples of spikes arising in a
European football match: a Bundesliga encounter between
Bayern Munich and Borussia Dortmund on May 4th, 2013.
The vertical axis
records the volume of Twitter activity over each one-minute period.
Here, a tweet is deemed to take part in the conversation if it
contains one or more specified keywords.
The spikes in bandwidth
can be attributed to unpredictable  external events, including goals and
controversial refereeing decisions, as indicated in the figure,
with a typical half-life of between ten and  twenty minutes.
Further Twitter spike examples are illustrated in the next section and in
[Supplementary Information].
These dramatic, but short-lived, bursts of interest represent marketing opportunities for suitably agile players, as demonstrated by the cookie company Oreo
who produced an effective, and, subsequently award winning, 
tweet in response to a power failure during Super Bowl XLVII
\cite{Post2013}.

Several authors have considered how
information is passed in the setting of online social media.
In \cite{Cha10} the number of followers, Retweets and mentions were used to quantify the influence of Twitter users,
with the three measures yielding very different results.
Similarly, \cite{Kwak10}
ranked users by the number of followers and also by Google's PageRank algorithm.
Related work in
\cite{Ler12}
looked at
how network structure affects dynamics of large scale information flow around news stories in
 Digg and Twitter.
In \cite{centola2010} the spread of behavior was examined through artificially constructed, static,
social interaction networks, with clustered-lattice structure found to be the most effective in
terms of speed and reach.
Dynamic analogues of the standard Katz centrality measures were tested on a large scale Twitter data set in
\cite{SNAM2013}, and found to be compatible with the rankings produced by
social media experts whose job is to identify key targets.

Looking more specifically at temporal patterns within online behavior,
\cite{Math2013}
used empirical long-time Twitter interaction data
to
show that the characteristic bursts of
activity are compatible with trading volumes of financial securities, and
proposed a stochastic point process model to reproduce the distribution of activity levels
across time.
Person-to-person
\emph{cascades}  of information spread have been studied by a number of authors
\cite{Bakshy11,Gleeson14,Leh2012,Leskov2007,Romero2011,Z13}; and we recommend
\cite{Borge2013} for an overview
of models and applications.

Our work differs from previous studies in three main respects.
First, rather than looking at the development of cascades \emph{within} a community (for example the rise of a viral video)
we focus on the ``event-driven, full attention span''
setting where the relevant community has been roused by an \emph{external} development,
as illustrated by the instances labelled 
in Figure~\ref{fig:spikes}.
In this type of spike phase, because  interest  levels are high, there is
a clear opportunity for targeted interventions to make an impact.
Second, we develop a model that addresses both
the dynamic nature of message-passing and the essentially static structure of the underlying
    \lq\lq who listens to whom\rq\rq\ network.
Third, by making our key modelling assumption explicit and developing a simple
algorithm that applies to large scale data sets, we produce a tool that can be employed in real time,
predicting who will be the most active players as soon as
a spike in volume has been detected.
Predictions from the new algorithm are tested on a Twitter data set, 
  with further tests reported  in [Supplementary Information].


\section{Method and Results}
\label{sec:method}

The big picture aim of this work is to understand what drives microblogging activity
during a \emph{full attention span} spike phase.
More specifically, we aim to develop an algorithm that could be used to monitor the network in real time,
and identify who would be the most active players if a spike were to flare up.
This forms the initial level of agility required to engage in real-time exploitation of the
raised awareness across the user base.

In setting up a general modelling framework, 
we assume that 
no new associations are created during the short time scale of the spike; that is,
we have a static underlying connectivity structure. 
To be concrete, we will discuss Twitter activity, but we note that the same principles apply to
other time-dependent digital messaging systems.
For the relevant set of $N$ users, we let $A \in \RR^{N \times N}$ denote a corresponding adjacency matrix
where $a_{ij} = 1$ if user $i$ is known to receive and take notice of messages from
user $j$. Loosely, this might mean that $i$ is known to be a Twitter follower of $j$, although in practice
we have in mind the use of more concrete evidence that $i$ cares about the tweets of $j$; for example
via Retweets.
For simplicity, we
take a standard unit of time
(one minute in the tests below)
during which a user is
assumed to send out at most one
message.
We let
$s^{[k]} \in \RR^{N}$ denote an indicator vector for the send activity at time $k$, so that
$s^{[k]}_i = 1$ if user $i$ tweeted in time interval $k$ and
$s^{[k]}_i = 0$ otherwise.
Then simple bookkeeping
tells us that
\begin{equation}
   r^{[k]} = A s^{[k]},
\label{eq:rs}
\end{equation}
where
$r^{[k]} \in \RR^{N}$  is such that
$r^{[k]}_i$ counts how many messages were received by user $i$ in this time interval.

We can now formalize our main modelling assumption. In words, the probability of
a user tweeting at time $k+1$ \emph{is proportional to the number of significant tweets
they have just received}, with proportionality constant denoted $\alpha$, plus a basal rate.
We therefore model $s^{[k]}$ as a discrete time Markov chain according to
\begin{equation}
  \PP \left( s^{[k+1]}_i  = 1 \ | \ s^{[k]} \right)  = b_i + \alpha \, r^{[k]}_i.
\label{eq:rs2}
\end{equation}
Here $b_i$ denotes the basal tweet rate for user $i$ and the second term on the right-hand side
quantifies our assumption that, in the full attention span phase, activity is driven
by a desire to join in with the current conversation and engage in topical \lq\lq banter.\rq\rq\ 
Formally,  
a normalization factor should be included in the right-hand side of
equation 
[\ref{eq:rs2}], to guarantee that probabilities lie between
zero and one. However, we will see that for our purpose of ranking nodes, this is not
necessary.

As general support for the key modelling assumption,
we note that
\cite{Bakshy2012}
found
social influence to play a crucial role in the propagation of information on
Facebook:
\lq\lq
Those who are exposed [to friends' information] are significantly more likely to spread information
and do so sooner than those who are not exposed.\rq\rq\
%
%
Further
empirical work appeared in \cite{LKML13}, which
looked at Twitter interactions under shared activity around eight major events during the
2012 U.S.\ presidential election.
The study found that human behavior changes during
a \lq\lq media activity,\rq\rq\ when information consumption is
characterized by the availability of dual screening technology (television and hand held device)
and real-time interaction.
The authors proposed the term
\emph{media event-driven behavioral change}
for this general effect, and showed that, for the data they collected,
differences in behavior were driven
by the increasing attention given to a small cohort of elite users.
Our work also focuses on  this shared-attention, event-based setting, and the
leadership role of
central users, but
considers behavior when the whole network rapidly becomes aware of
an item of breaking news.


Combining 
equations
[\ref{eq:rs}]
and 
[\ref{eq:rs2}],
we see that
\begin{equation}
  \PP \left( s^{[k+1]}_i = 1  \ | \ s^{[k]} \right)  = b_i + \alpha \, \left(  A s^{[k]} \right)_i.
\label{eq:probdyn}
\end{equation}
It follows that the expected value
$\EE[  s^{[k+1]} ]$ evolves according to
\begin{equation}
\EE [
 s^{[k+1]} ]
     = b + \alpha A \EE[ s^{[k]} ];
\label{eq:iter}
\end{equation}
see [Supplementary Information].
This type of iteration is familiar in many modelling and computation scenarios,
notably in numerical analysis,
and it is readily shown that
as $k$ increases $\EE[ s^{[k+1]} ]$ generically lines up
along a preferred direction that is independent of $s^{[0]}$; see
\cite[Theorem~10.1.1]{GVLbook}
for details, or 
[Supplementary Information]
for an informal treatment.
If the spectral radius of $A$ is below $1/\alpha$
then as  $ k \to \infty$ the resulting steady state
value for $\EE [ s^{[k]} ]$, which we denote by
$s^\star$, satisfies
\begin{equation}
(I - \alpha A ) s^\star = b.
\label{eq:sstar}
\end{equation}

Overall,
having constructed the matrix $A$ and the right-hand side $b$ from the current data,
 the vector $s^\star$ in equation [\ref{eq:sstar}] can be used to predict the typical activity
level of each node in the event of a spike, a larger value of $s^\star_i$ suggesting that user $i$ will
be more active.
In particular, the current top $r$ components in the
vector $s^\star$ give a prediction for who would be the $r$ most active users in the event of
a spike.

To compute $s^\star$ in equation [\ref{eq:sstar}] requires the solution of a
linear system involving the underlying network adjacency matrix, $A$.
Because $A$ is typically very sparse (i.e., has only a small percentage of nonzeroes),
this computation is feasible; for example, on a typical current desktop machine the number of nodes
can be in the millions.
We may regard $s^\star$ as a \emph{network centrality measure} \cite{Newmanbook};
indeed it is closely related to the widely-used Katz centrality
\cite{Katz53}, and can be
interpreted independently from a combinatoric, graph-theoretic standpoint; see
[Supplementary Information].
In the special case where $\alpha = 0$, we do not make use of any underlying network
information, and predict purely on the
basal rate of each user.
This provides a natural
basis for testing the algorithm, and therefore
validating our underlying hypothesis: does the use of
$\alpha > 0$
add value to the prediction of who will be active during a spike?

We address this question using a Twitter data set,
with three further
sets tested in
[Supplementary Information].
In each case, we define a \emph{business as usual}
period where users operate at their basal rate
and a \emph{spike period},
where
network activity has been dramatically increased by an external event.
The basal tweet rate $b_i$ for user $i$ is taken to be their total number of
business as usual period tweets.
As mentioned above,
since we are only concerned with the relative ranking induced by $s^\star$
in equation [\ref{eq:sstar}],
there is no requirement to normalize this quantity.
We also build the matrix $A$ from business as usual data,
setting $a_{ij} = 1$ if $i$ received at least one relevant tweet from $j$ in this period, and
setting $a_{ij} = 0$ otherwise.
We therefore wish to judge the
predictive power of $s^\star$
during the spike period.
We do this by predicting key users
with  $s^\star$ and
 recording their total Twitter activity during the spike period.

Our experiment uses data collected on May 9th 2013 surrounding the appointment
of David Moyes as manager of Manchester United Football Club, following the
retirement of Sir Alex Ferguson. This consisted of
298,335
 time-stamped directed message-passing events
involving
148,918
distinct Twitter accounts.
The upper picture in
Figure~\ref{Fig:mufcresp} shows the
volume of tweets each minute.
The largest spike in volume, at 486 minutes, corresponds to the official
announcement of Moyes' appointment.
(The next largest peak, appearing earlier, corresponds to
Everton Football Club announcing Moyes' departure.)
For the purposes of our test, we regard zero to  300
minutes as forming the business as usual
period where users operate at their basal rate.
We define the spike period as lasting from the peak time of 486 minutes to the time
of 541 minutes at which the
activity level has decayed by a factor of four.

As support for our modelling hypothesis that, in a spike phase,
activity is driven by a desire to engage with incoming messages,
we show in the lower picture of
Figure~\ref{Fig:mufcresp}  the \emph{responsiveness} of the network, defined as the
number of tweets that a typical sender has seen in the previous one minute of their timeline.
More precisely, 
we compute the
responsiveness over the $i$th one minute period as
\begin{equation}
 \frac{1}{N_i}
 \sum_{k = 1}^{N_i}  \mathrm{rec}_i^{[k]},
 \label{eq:respdef}
\end{equation}
where $N_i$ denotes the number of tweets sent out in this minute and, for each such tweet,
$ \mathrm{rec}_i^{[k]}$ denotes the number of tweets that the sender received in the
previous 60 seconds.

Now, we test the predictive power of the new measure
$s^\star $ in equation [\ref{eq:sstar}] as a function of the response rate parameter, $\alpha$.
Figure~\ref{Fig:mufclog} shows the change in 
\emph{total spike period activity of the top 100 ranked users},
 as a function of $\alpha$.
In other words, for each choice of $\alpha$ we use the
business as usual information to compute $s^\star $, find the
users with the 100 top-ranked values of $s^\star $ and then
record the total number of tweets sent by this top 100 during the spike period.
The figure shows the difference between the total activity of these users and those 
from the baseline value of 
$\alpha = 0$.
For compatibility with the other three tests reported in 
[Supplementary Information],
 we present the results in terms of
the normalized parameter $\alpha^\star = \alpha/\rho(A)$,
where $\rho(\cdot)$ denotes the spectral radius, so that
$\alpha^\star = 1$ becomes a natural upper limit.
The $\alpha^\star = 0$ baseline is marked with a dashed line.
In this example,  as soon as $\alpha^\star $ increases beyond machine precision level
(around ${10}^{-16}$), the top 100 list changes and the prediction improves.
We  also see that there is  a broad range of $\alpha^\star$ values for which
an improved prediction is obtained, relative to the
$\alpha^\star = 0$ case where no underlying connectivity is exploited.

We emphasize that this test was designed to use only data available in the
business as usual phase in order to predict activity in a subsequent spike phase.
In 
[Supplementary Information]
we present three further case studies involving
a sporting event, a marketing event and a TV programme.
On the basis of these tests, we conclude that there is
value to be had from folding in information about the underlying static network structure:
in each case,
incorporating a small value of the response  parameter, $\alpha$, does not degrade the
prediction, and leads to an improvement for a broad range of choices.

We also note that our new  model can be used to explain the
characteristic geometric decay in tweet volume
observed in these examples
following a peak of activity.
In particular, the half-life of a spike can be estimated as
\begin{equation}
  \left|
 \frac{\log{2}}{\log(\gamma)} \right|,
\label{eq:hl}
\end{equation}
where  $\gamma$ is the product of two factors: the response rate $\alpha$ and the
Perron--Frobenius eigenvalue  of $A$. The latter may be regarded as roughly the maximum
number of followers over all relevant users, and is hence a measure of community size.
See 
[Supplementary Information]
for further details.

\section{Discussion}
\label{sec:disc}
This work tackled the important setting where there is a spike in social interaction caused by an external
event.  We took the novel step of incorporating both the static, underlying
\lq\lq who knows whom\rq\rq\
network and the dynamic \lq\lq who is currently active\rq\rq\ information.
By making a quantifiable hypothesis that, in this special phase, our activity increases if we see our friends becoming involved in the conversation,  we were able to develop a testable algorithm
that predicts activity levels in the event of spike. This type of information is
of great value to those wishing to control, or interfere with, the rapid spread of
information 
during a spike.
The new algorithm  has been validated on data from Twitter conversations around high profile events.
However, we emphasize that the underlying concepts  are relevant to any other
digital social media setting where we pass information in real time to a pre-specified
group of social neighbors.

Further work in this direction will include (a) testing the algorithm on other
data sets from a range of social media settings and
(b) looking at optimal methods for constructing the static interaction matrix $A$, the
basal activity vector, $b$ and the response strength parameter, $\alpha$.

\begin{figure}
\begin{center}
 \includegraphics[height=100mm]{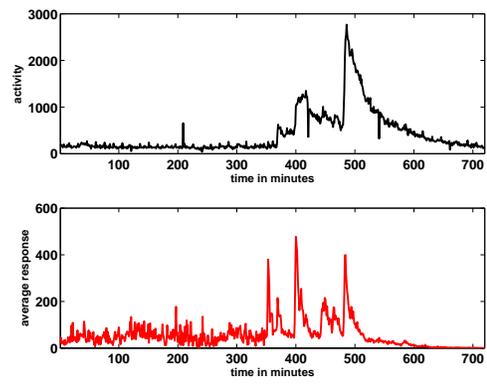}
\caption{
Upper: volume of tweets each minute
for a conversation around Manchester United Football Club, May 9th, 2013.
Lower:
 responsiveness of the network, defined in equation [\ref{eq:respdef}].
}
\label{Fig:mufcresp}
\end{center}
\end{figure}

\begin{figure}
\begin{center}
 \includegraphics[height=100mm]{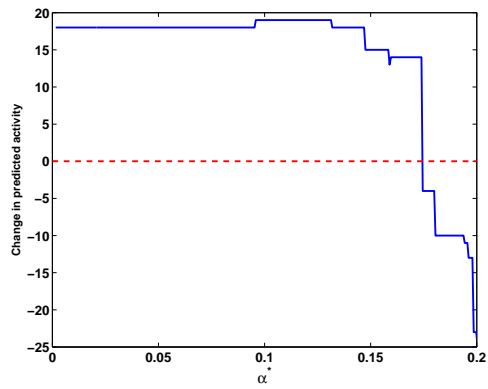}
\caption{
Solid line:
change in activity of predicted top 100 tweeters during the spike phase in
Figure~\ref{Fig:mufcresp}, as a function of the response
parameter, $\alpha^\star$.
Dashed line: corresponding level for $\alpha^\star = 0$.
}
\label{Fig:mufclog}
\end{center}
\end{figure}

\bigskip
\noindent
\textbf{Acknowledgements}\\
We are grateful to Bloom Agency, Leeds, UK, for supplying the Twitter data.
PG was supported by the Research Councils UK Digital Economy Programme via
EPSRC grant
EP/G065802/1 \emph{The Horizon Digital Economy Hub}.
DJH acknowledges support from a Royal Society Wolfson Award and a Royal Society/Leverhulme
Senior Fellowship.
AVM was supported by the EPSRC and Bloom Agency via an Impact Acceleration Account secondment.
Data was acquired using a grant from the University of Strathclyde through their
EPSRC-funded \lq\lq Developing Leaders\rq\rq programme.
The Manchester United data set used for 
Figures~\ref{Fig:mufcresp} and \ref{Fig:mufclog} will be made publicly 
available soon at the URL\\ \verb5http://www.mathstat.strath.ac.uk/outreach/twitter/mufc5





\begin{figure}
\vspace{.025 cm}
\begin{center}
 \includegraphics[height=50mm]{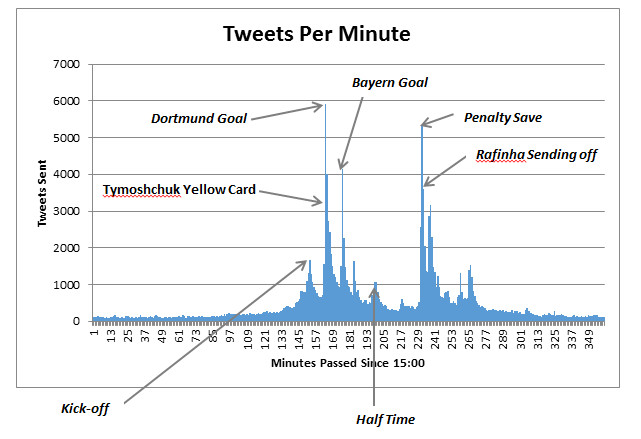}
\caption{
Tweets in each one minute interval concerning a
European
football  match.
}
\label{fig:spikes}
\end{center}
\end{figure}

\clearpage

\section{Supplementary Information}
\label{sec:supp}

\subsection{Evolution of the Expected Value}
\label{sec:exp}

We note that because $s^{[k+1]}_{i} $ takes only  the value $0$ or $1$, 
its conditional expectation is simply the probability of taking the value $1$, given
$s^{[k]}$. So, taking conditional expectation in 
equation [\ref{eq:probdyn}], we have 
\[
\EE [
 s^{[k+1]}_i | s^{[k]} ]
     = b_i + \alpha  \left( A s^{[k]} \right)_{i}.
\]
Upon taking expected values, we then obtain 
\[
\EE [
 s^{[k+1]}_i ]
     = b_i + \alpha  \left( A  \EE [ s^{[k]} ] \right)_{i},
\]
giving 
equation 
[\ref{eq:iter}].

\subsection{Stationary Iteration}
\label{sec:iter}

For the iteration in equation [\ref{eq:iter}], we have 
\begin{eqnarray*}
\EE[ 
 s^{[1]} ]
     &=& b + \alpha A \EE [  s^{[0]} ] \\
\EE[  s^{[2]} ]  &=& b + \alpha A \left( b + \alpha A \EE [  s^{[0]} ] \right)
 =  b + \alpha A  b + (\alpha A)^2  \EE [  s^{[0]} ],
\end{eqnarray*}
and the general pattern, which may be  
proved formally by induction, is 
\[
\EE[ 
s^{[n]} 
] = b + \alpha A  b + \cdots +  (\alpha A)^{n-1}  b + 
                          (\alpha A)^{n}  \EE[ s^{[0]} ].
\]
Under our assumption that 
$\alpha < 1/\rho(A)$, 
it follows that $\|  (\alpha A)^{n} \| \to 0$ as $ n \to \infty$, for
any matrix norm 
$\|  \cdot \|$.
 Hence, the influence of $s^{[0]}$ becomes negligible, and 
$
\EE[ 
 s^{[n]} 
]
$
approaches 
\[
  \sum_{i = 0}^{\infty}  (\alpha A)^{i}  b,
\]
which may be written 
$(I - \alpha A)^{-1} b$.

\subsection{Katz-like parameter}
In our notation, where $A$ denotes an 
adjacency matrix, the $k$th power, $A^k$, has an $(i,j)$ element that 
counts the number of directed walks 
from node $i$ to node $j$.
It follows that the infinite series
\[
I + \alpha A + \alpha^2 A^2 + \cdots + \alpha^k A^k  + \cdots 
\]
has $(i,j)$ element that 
counts the total number of walks from 
from node $i$ to node $j$ of all lengths, where 
a walk of length $k$ is scaled by $\alpha^k$.
Here \lq\lq length\rq\rq\  refers to 
the number of edges traversed during the walk.
This series converges 
for 
$0 < \alpha < 1/\rho(A) $, whence it may be written 
$(I - \alpha A)^{-1}$.

The vector $c \in \RR^{N}$
defined by
$ c = (I - \alpha A)^{-1}  \mathbf{1}$, or, equivalently,
\[
 (I - \alpha A) c =  \mathbf{1},
\]
where 
 $\mathbf{1} \in \RR^{N}$ denotes the vector with all values equal to unity, 
 therefore has $i$th element that 
 counts the number of directed walks 
  from node $i$ to every node in the network, with 
 a walk of length $k$ scaled by $\alpha^k$.
This is one way to measure the \lq \lq centrality\rq\rq\ of node $i$, as first 
 proposed by Katz 
\cite{Katz53}.
In this way, $\alpha$ becomes the traditional attenuation parameter in the Katz setting, representing 
the probability that a message successfully traverses an edge. 
The measure 
$
 s^\star 
$ 
 in 
equation 
[\ref{eq:sstar}]
replaces the uniform vector
$
\mathbf{1}
$
with $b$.
Hence,
the component $s^\star_i$ can be 
interpreted as a count of the total number of walks from node $i$ to every node in the network, with walks to node $j$ weighted by $b_j \alpha^k$. 
The introduction of $b$ has therefore allowed us to weight the walk count  
according to  basal dynamic activity.

\subsection{Half-Life of a Spike}
\label{subsec:half}

At the start of a spike, it is reasonable to suppose that 
$\EE[ s^{[0]}]$ in 
equation 
[\ref{eq:iter}]
is very large. We then have  
\[
\EE[ s^{[1]}] = b + \alpha A  \EE[ s^{[0]}] \approx \alpha A  \EE[ s^{[0]}],
\]
and generally, in this spike phase, 
\begin{equation}
\EE[ s^{[k]}] \approx (\alpha A)^k  \EE[ s^{[0]} ].
\label{eq:kpow}
\end{equation}
In the regime where 
$\alpha \rho(A) < 1$ it follows that the expected level of activity decays over time.
More precisely, if we assume that the nonnegative matrix $A$ is
irreducible (that is, 
every node in the network has a path to every other) then the Perron--Frobenius Theorem
\cite{HigFM} says that there is a unique, real, positive, largest eigenvalue, $\lambda_1$ with 
corresponding 
nonnegative
eigenvector 
$v_1$.
We will expand $\EE[ s^{[0]} ]$ as 
$\sum_{i=1}^{N} \beta_i v_i$, where $\{ v_i \}_{i=1}^{N}$ are the 
eigenvectors of $A$, which we assume to span $\RR^{N}$, 
with corresponding eigenvalues 
$\{ v_i \}_{i=1}^{N}$ and with 
$\beta_1 > 0$.
Then in equation [\ref{eq:kpow}], 
\[
    \EE[ s^{[k]}] \approx  \sum_{i=1}^{N} 
           \beta_i (\alpha \lambda_i)^k   v_i.
\]
Since $\lambda_1$ is dominant, we have 
\[
    \EE[ s^{[k]}] \approx  
           \beta_1 (\alpha \lambda_1)^k   v_1, 
\]
so
\[
\mathbf{1}^T 
    \EE[ s^{[k]}] \approx  
           \beta_1 (\alpha \lambda_1)^k  
\mathbf{1}^T v_1.
\]
We conclude that 
$\mathbf{1}^T
    \EE[ s^{[k]}]$,
the overall expected network activity 
at time $k$, 
satisfies
\[
\mathbf{1}^T 
    \EE[ s^{[k]}] \approx  
               C 
            (\alpha \lambda_1)^k, 
\]
where $C$ is a constant independent 
of $k$. 
The half-life then corresponds to $\widehat k$ time units, where 
\[
 (\alpha \lambda_1)^{\widehat k} = \frac{1}{2},
\]
leading to the expression in equation [\ref{eq:hl}].

The 
Perron--Frobenius eigenvalue, $\lambda_1$,
is bounded 
above by 
any subordinate matrix norm. 
Taking the standard 
$\| \cdot \|_{1}$ or
$\| \cdot \|_{\infty}$ corresponds to forming 
the maximum in-degree or out-degree, respectively.

\subsection{Further Twitter Case Studies}
\label{subsec:case}

\begin{figure}
\vspace{.025 cm}
\begin{center}
 \includegraphics[height=100mm]{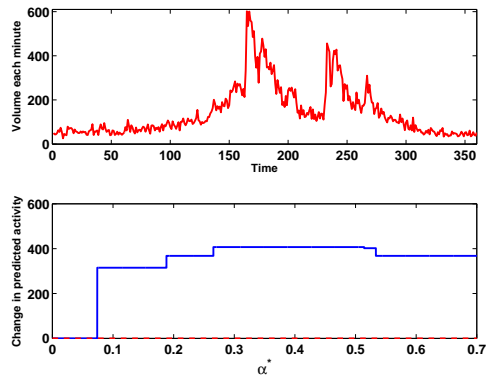}
\caption{
Upper: volume of tweets each minute 
for a conversation around a football match. 
Lower: change in activity of predicted top 10 tweeters during the spike phase, as a function of the response 
parameter, $\alpha^\star$.
Dashed line: corresponds to $\alpha^\star = 0$.
}
\label{Fig:bayernfig}
\end{center}
\end{figure}

Figure~\ref{Fig:bayernfig}
presents results  
for the football match
data shown in Figure~\ref{fig:spikes}.
This involves 37,479  Twitter users. 
The upper picture shows Twitter volume per minute.
We regard time zero to 130 minutes as the business as usual period,
and define the spike period as starting at the peak of   
165 minutes and finishing at 175 minutes, after which activity starts to increase again.
This data is an order of magnitude smaller that the Manchester United 
data in Figure~\ref{Fig:mufcresp},
so we focused on the predicted top 10 users.
The lower picture in
Figure~\ref{Fig:bayernfig}
shows the 
change in 
total spike period activity of this top ten as a function of $\alpha^\star$. 

\begin{figure}
\vspace{.025 cm}
\begin{center}
 \includegraphics[height=100mm]{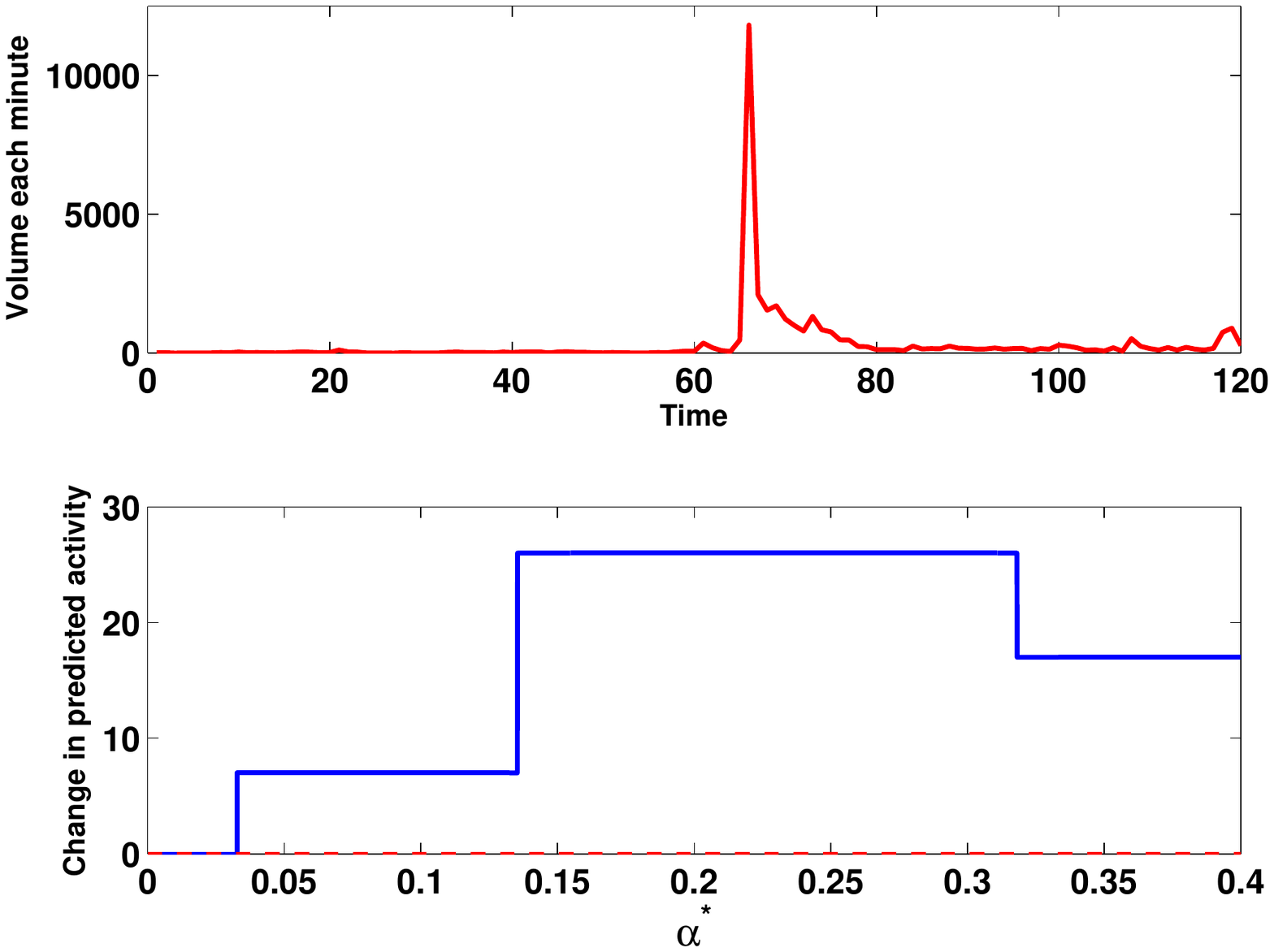}
\caption{
Upper: volume of tweets each minute 
for a conversation around a marketing event. 
Lower: change in activity of predicted top 10 tweeters during the spike phase, as a function of the response 
parameter, $\alpha^\star$.
Dashed line corresponds to $\alpha^\star = 0$.
}
\label{Fig:teafig}
\end{center}
\end{figure}

For Figure~\ref{Fig:teafig}, we used data from a marketing event 
for the Yorkshire Tea Company on April 24th, 2013, where a range of 
tea lovers and celebrities, including Louis Tomlinson from pop band One Direction,  
took part in an Orient-Express style train journey around Yorkshire, UK, and were encouraged to 
publicize the event.
In this case we have 
9,163 Twitter users.
The large spike at  66 minutes corresponds to awareness being raised about the presence 
of a One Direction member. 
We defined the business as usual period to last from zero to 65 minutes.
The lower picture 
shows the change in spike
activity of the predicted top ten as a function of 
the response parameter $\alpha^\star$.

\begin{figure}
\vspace{.025 cm}
\begin{center}
 \includegraphics[height=100mm]{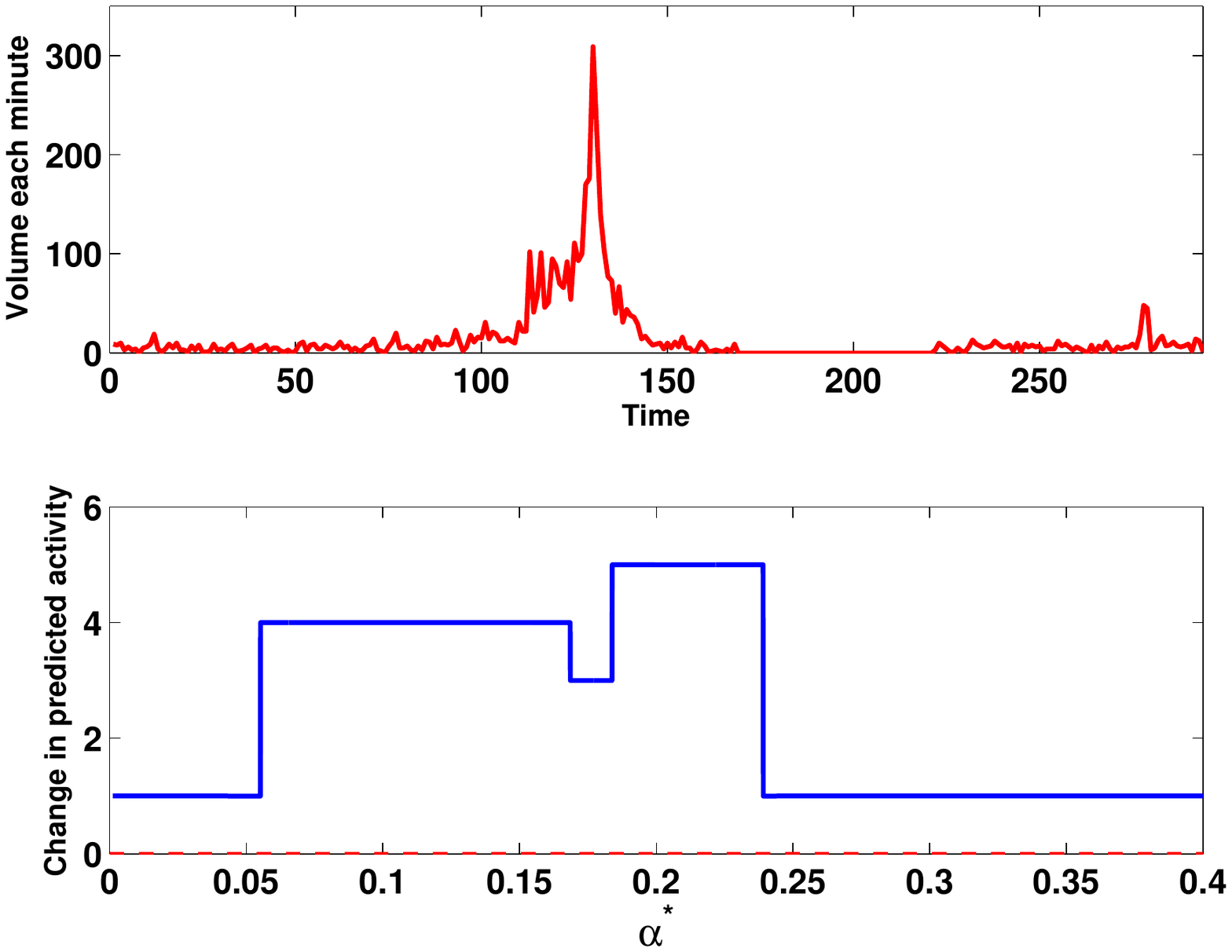}
\caption{
Upper: volume of tweets each minute 
for a conversation around a TV programme. 
Lower: change in activity of predicted top 10 tweeters during the spike phase, as a function of the response 
parameter, $\alpha^\star$.
Dashed line corresponds to $\alpha^\star = 0$.
}
\label{Fig:utopiafig}
\end{center}
\end{figure}

Figure~\ref{Fig:utopiafig}
shows results for a dual--screening conversation
around an episode of the Channel Four UK television programme 
Utopia, involving
4,154 Twitter users.
The spike at time 130 minutes corresponds to a particularly dramatic scene.
We defined the spike to finish at 145 minutes, and took the business as usual period to last from time zero to 120 minutes.
As before, the change in  
spike activity as a function of $\alpha^\star$ is shown in the lower picture.  

In each of these three further tests, we see that 
extra value is added by 
increasing $\alpha^\star$ above zero; that is, by 
appropriately incorporating 
information about the 
underlying follower network 
that was built up 
in 
advance of the spike.

\end{document}